\documentclass[a4paper]{ESASPCS13Style}
\usepackage{epsfig}

\begin{document}

\title{Binaries among Ultra-cool Dwarfs and Brown Dwarfs}

\author{Wolfgang Brandner\inst{1}, Herv\'e Bouy\inst{2}, Eduardo L.\ Mart\'{\i}n\inst{3} and Stefan Umbreit\inst{1}} 
\institute{Max-Planck-Institut f\"ur Astronomie, K\"onigstuhl 17, D-69117 Heidelberg, Germany
  \and Max-Planck-Institut f\"ur Extraterrestrische Physik, Giessenbachstrasse, D-85748 Garching, Germany
\and Instituto Astrof\'{\i}sica de Canarias, E-38200 La Laguna, 
             Tenerife, Spain}

\maketitle 

\begin{abstract}

Observations of brown dwarfs provide important feedback on 
theories of atmospheres and inner structure of substellar objects. 
Brown dwarf binary systems furthermore offer the unique opportunity to 
determine the mass of individual brown dwarfs, which is one of the fundamental 
astrophysical quantities.

\end{abstract}

\section{Introduction}
  
Since brown dwarfs only briefly stabilize themselves on a Deuterium
burning ``main-sequence'', after which they continue to contract and
to cool down, there is in general a degeneracy between brown dwarf
mass and luminosity or age and temperature. Dynamical mass estimates
derived from the orbital parameters of binary brown dwarfs are the 
only means to measure one of the fundamental astrophysical quantities,
and to improve our understanding of substellar objects. Binary properties
of brown dwarfs should reflect on their formation mechanism, and hence
provide insights into the origin of brown dwarfs. Finally, since brown
dwarfs are considerably less luminous than stars, it is much easier to
identify young, still self luminous planetary mass objects in orbit around
brown dwarfs than around stars.

\section{The Sample}

 \begin{figure}[htb]
\centerline{
 \psfig{figure=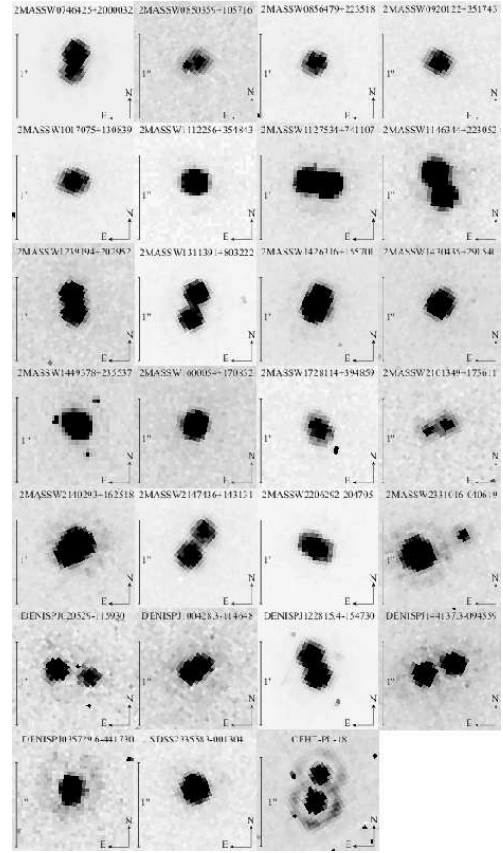,width=6.5cm,angle=0}}
\caption{\label{fig1} Nearby binary brown dwarfs and ultra-cool dwarfs
with separations between 1 and 20\,A.U.\ (Bouy et al.\ 2003)}
\end{figure}

Our group has been studying binary brown dwarfs
since 1998. In the following we summarize ongoing surveys for spatially 
resolved brown 
dwarf binaries, and present results of follow-up studies using HST and 
ground-based adaptive optics.

More than 140 nearby ultra-cool dwarfs and brown dwarfs with
spectral types late M (later than M7) and L have by now been surveyed for 
companions by Close et al.\ (2002a, 2002b) using ground-based adaptive optics
(Hokupa$'$a) at Gemini North, and by Reid et al.\ (2001), Bouy et al.\ (2003),
Gizis et al.\ (2003) and Golimowski et al.\ (2004) using the Hubble Space 
Telescope (HST).

As the typical separation of the binary components is less than 0.5$''$,
high-angular resolution observations are mandatory if one wants to study
the properties of the individual components of these binary systems.

The 27 binaries studied in greater detail by Bouy et al.\ (2003)
are shown in Figure \ref{fig1}.

\section{Brightness Ratios and Semi-major Axes}

\begin{figure}[htb]
\psfig{figure=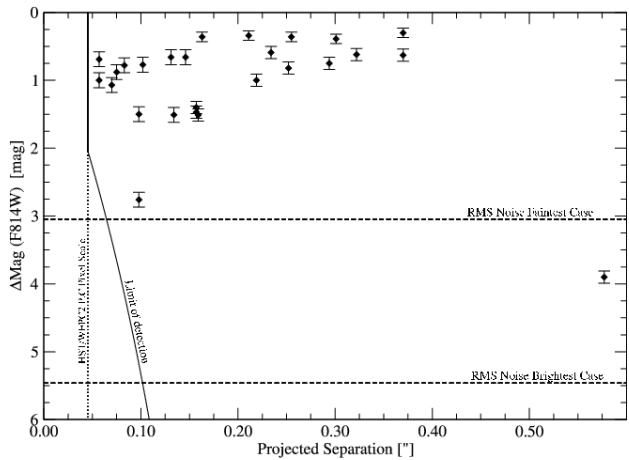,width=6.0cm,angle=270}
\caption{\label{fig2} In most cases, the secondaries are
only slightly redder and fainter than the primaries, suggesting
a preference for
equal-mass systems and a lack of systems with extreme mass ratios.}
\end{figure}

The parameter space covered in brightness ratio ($\Delta$\,mag) and separation
is shown in Figure \ref{fig2}. Even though it would have been
possible to detect companions up to 5.5\,mag fainter than the primaries,
the observed brightness differences are in general $\le$1\,mag. This indicates
a strong preference for equal mass binaries among brown dwarfs.

\begin{figure}[htb]
 \psfig{figure=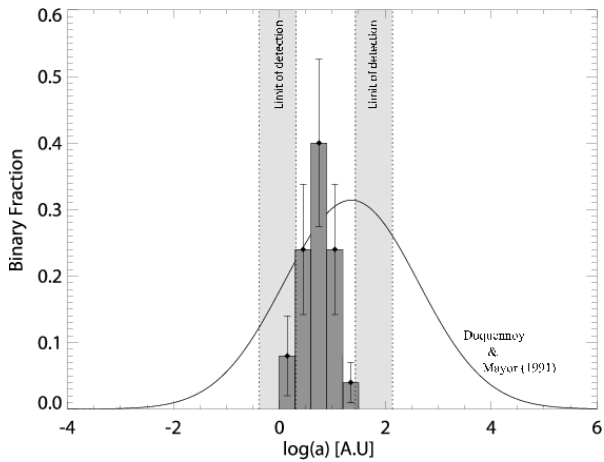,width=6.0cm,angle=270}
\caption{\label{fig3} The distribution of semi-major axis is clearly
distinct  from G-type binaries according  to Duquennoy \& Mayor (1991) -- in 
particular, there are no wide
binaries among the ultra-cool dwarfs and brown dwarfs.
}
\end{figure}

Figure \ref{fig3} compares the distribution of semi-major axis
to the distribution found among solar-type stars. The most striking
feature is the lack of wide binaries with separations larger than 
$\approx$20\,A.U.\ among the brown dwarfs. The lack of brown dwarf
binaries with separations smaller than $\approx$1\,A.U.\ might still
be a detection bias, since such systems would be unresolved even at
the angular resolution provided by HST.

\section{Orbits \& Masses}

\begin{figure}
\vbox{
\centerline{\psfig{figure=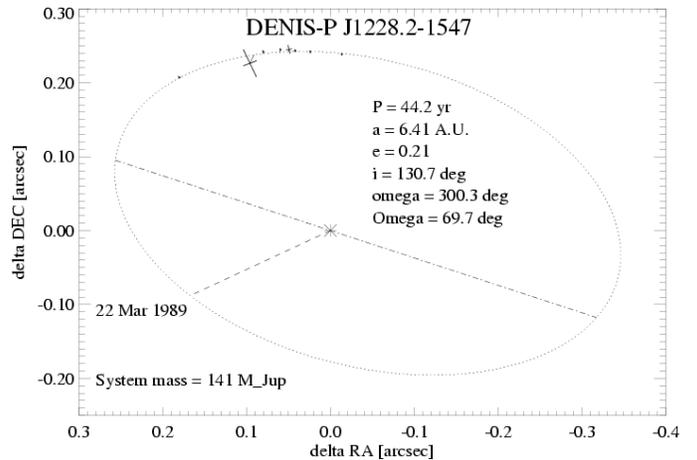,width=6.0cm,angle=270}}
\centerline{ \psfig{figure=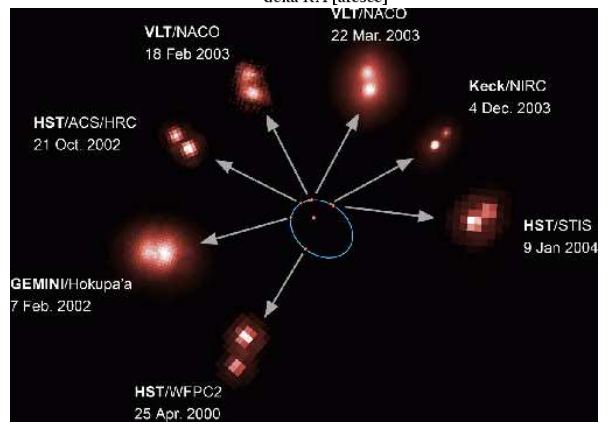,width=5.5cm,angle=270}}
}

\caption{\label{fig4} Observations of the orbital motion of the
L-dwarf binaries DENIS-P\,J1228.2-1547 (top, Brandner et al.\ 2003) and
2MASSW\,J0746425+2000321 (bottom, Bouy et al.\ 2004) yield dynamical mass
estimates.}
\end{figure}

Once companions have been identified, second epoch observations
are important to confirm that the objects form indeed physical
binaries, and in order to start the monitoring of the orbital motion
of the two binary components around their common center of mass.
In Figure \ref{fig4} we show two examples for multi-epoch monitoring
of binary L-dwarfs with the aim to derive the orbital parameters,
and dynamical system masses.  On the top we show the results of 
5.5\,years of continued observations of DENIS-P\,J1228.2-1547, the
first spatially resolved binary L-dwarf (Mart\'{\i}n et al.\ 1999;
Brandner et al.\ 2003, 2004). On the bottom we show the results of
4 years of monitoring of 2MASSW\,J0746425 +2000321, the first L-dwarf
with a complete determination of the orbital parameters, and hence
the first system for which precise dynamical mass estimates have
been derived (Bouy et al.\ 2004). 

Monitoring programmes for other binary L- and T-dwarfs are on-going, and we
expect to derive more dynamical mass estimates within the next 5 to 10\,yr.

\section{Formation of Brown Dwarfs}

\begin{figure}[htb]
\vbox{
\psfig{figure=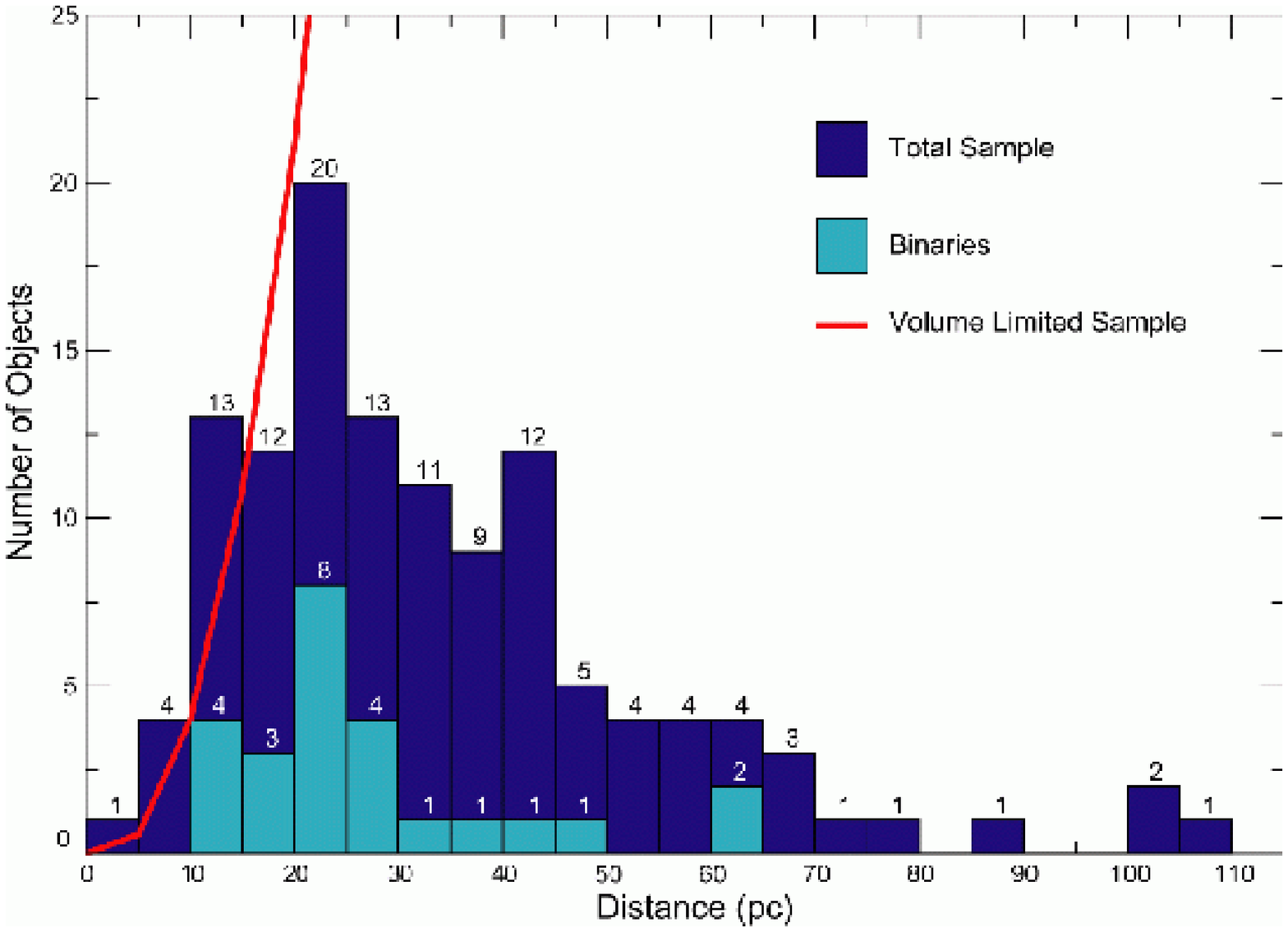,width=6.0cm,angle=270}
 \psfig{figure=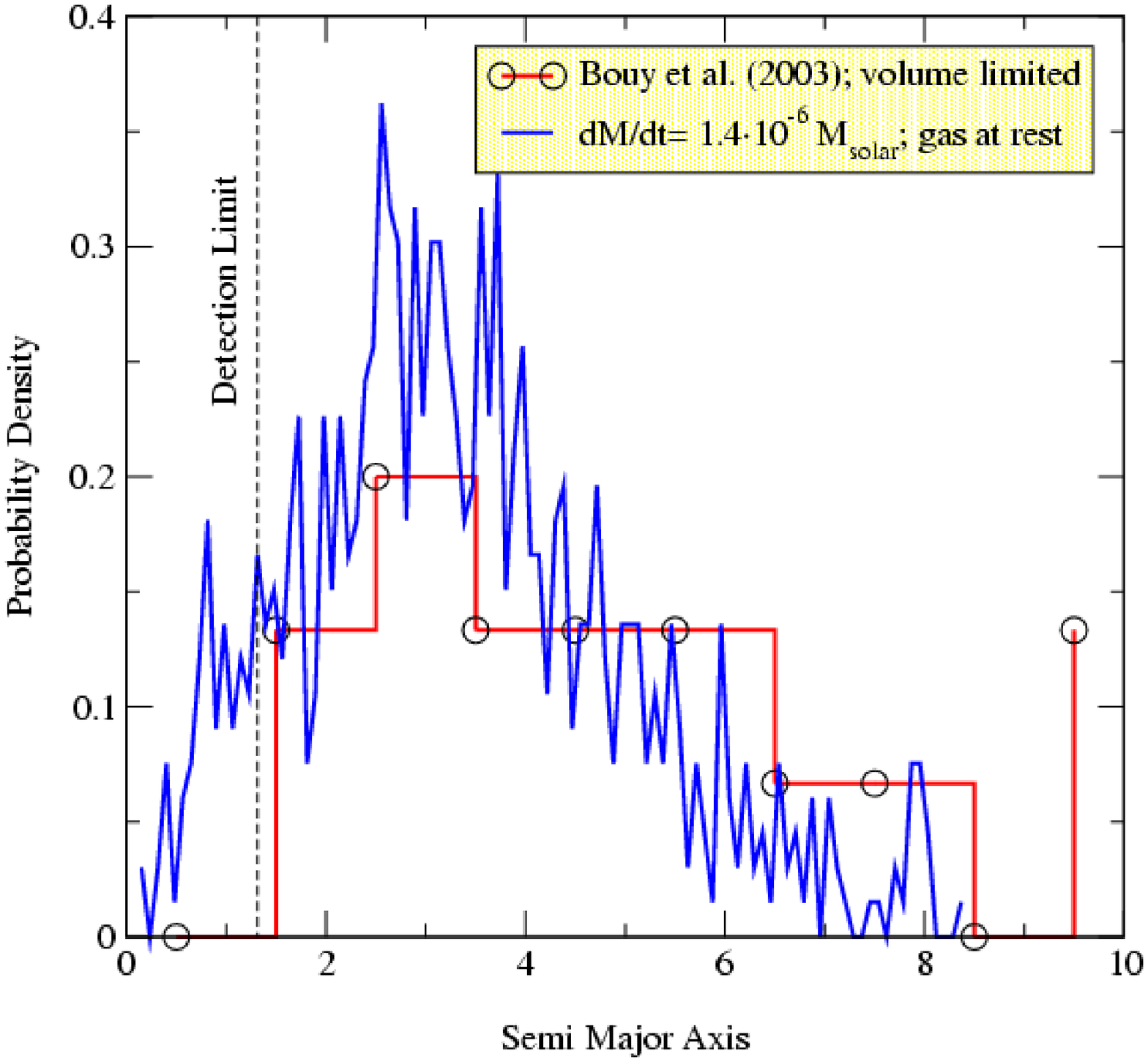,width=6.0cm,angle=270}
}
\caption{\label{fig5} Distance estimates allow for the transformation of the
magnitude limited sample (biased in favour of binaries) into a distance
limited sample (top). Simulations of brown dwarf binary formation
based on the Reipurth \& Clarke model (2001) show a remarkably good agreement
with the observed distribution of semi-major axis in the distance limited
sample (bottom, Umbreit et al.\ 2004).
}
\end{figure}

Apart from the determination of dynamical mass estimates and related physical
properties of individual brown dwarfs, our study also aims at a better
understanding of the formation processes of (binary) brown dwarfs.
One of the currently discussed formation processes is the so-called 
``embryo-ejection'' model (Reipurth \& Clarke 2001), where brown
dwarfs are ejected from multiple systems during the main-accretion
phase. Parameters studies of the dynamical interactions of three 
40\,M$_{\rm Jup}$, still accreting proto-brown dwarfs have been carried
out by Umbreit et al.\ (2004).

For comparison of the resulting separation distribution with observations,
our magnitude limited sample of binary L-dwarfs has first to be transformed
into a distance limited sample. As can be seen from Figure \ref{fig5}, top,
our sample is complete only out to distances of $\approx$22\,pc, hence
only binaries within 22\,pc are considered in the following. The
resulting distribution of binary separations for the distance limited
sample is shown as a histogram in Figure \ref{fig5}, bottom. Overplotted
in blue (continuous line) is the distribution of separations derived in the
simulations for one particular value of accretion rates. The good agreement 
between simulations and observations is very encouraging, though more detailed
simulations, and an extension of observational studies towards smaller
binary separations are required before any firm conclusions can be
drawn.

\section{The Next Step}

\begin{figure}[htb]
\psfig{figure=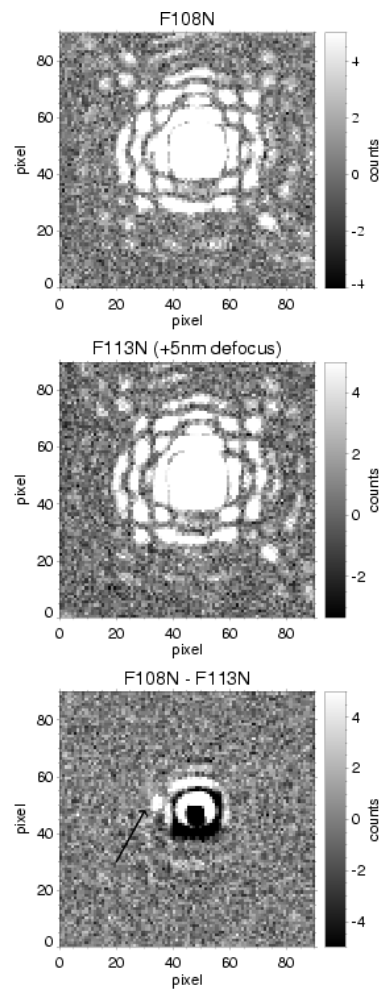,width=5.5cm,angle=180}
 \caption{\label{fig6} In HST Cycle 13, we target the twelve most nearby 
(within 30\,pc),
isolated (no known close companions), and young ($<$\,1\,Gyr) brown dwarfs
to search for planetary mass companions. By differential observations in and
off a molecular absorption band, the survey should be sensitive enough
to detect companions with masses as low as 6\,M$_{\rm Jup}$ and at separations
$>$\,3\,A.U.\ from the brown dwarf primary.}
\end{figure}

The Next Step is a survey for planetary mass companions to brown dwarfs.
Ongoing observations with HST aim at studying 12 of the most nearby and
young brown dwarfs for planetary mass companions. At an age of $\le$\,1\,Gyr,
objects with masses in the range 5 to 10\,M$_{\rm Jup}$ should still be
bright (hot) enough to be detectable by their intrinsic radiation.
In Figure \ref{fig6} we show simulations for observations with HST/NIC1
of a 6\,M$_{\rm Jup}$ companion to a brown dwarf at two different
wavelengths in and out of a molecular absorption band. The planetary
mass companion is nicely detected in the difference image shown on the right.

\end{document}